# Viewing galaxies in 3D

Davor Krajnović[1]

*Thanks to a technique that reveals galaxies in 3D, astronomers can now show that many galaxies have been wrongly classified. Davor Krajnović argues that the classification scheme proposed 85 years ago by Edwin Hubble now needs to be revised. [Based on the article published in Physic World, November 2011, Vol 24. No.11 (http://dx.doi.org/10.1088/2058-7058/24/11/43)]*

Ten-year-olds are great at galaxy formation. On a recent visit to an elementary school, I told a bunch of pupils that big galaxies in the Universe come in two types - some look like pancakes, the others like rugby balls - and that they often collide. Then I asked them to imagine what would happen if two galaxies that look like pancakes collide. The answer came fast and clear: "They would make a rugby ball." Okay, they were listening. "Correct! Sometimes this really does happen. But how would you make a rugby ball from two pancakes?" Silence. Some behind-the-ear scratching in the second row was followed by a few long gazes through the window at the playground, while in the front row shoe laces that were undone were quickly tied up. Then suddenly a hand shoot up. I raised my eyebrow unleashing excitement rarely seen even in a science meeting: "If there is a bit of jam on the pancakes," says a pupil, vividly showing with her fingers how they would look, "when they smash onto each other they will stick and fold and twist and become like a rugby ball".

Unfortunately, galaxies are not made of jam, but some do look like pancakes - thin discs of stars, gas and dust. And others do like rugby balls - stellar agglomerates with ellipsoidal shapes. They interact in many ways, but the most spectacular are cosmic colisions of galaxies. Understanding how galaxies are formed, how they transform between each other, and what physics is involved is a key topic in the field of observational cosmology.

Unravelling this mystery is relevant to the wider physics community too, because the luminous parts of galaxies that we see seem to be embedded in dark-matter halos, they contain massive black holes, and their interactions are striking (and huge) hydrodynamical laboratories. To advance our understanding one would like to introduce some order in the zoo of all possible galaxies that are observed; one would like to classify galaxies and learn something from it.

**Galaxy classification**

In astro-speech, pancakes are called discs, and they often have prominent spiral features, hence they are usually called *spirals*. Ellipsoidal rugby balls are, unsurprisingly, called *ellipticals*. But there are also mixtures, such as spirals with ellipsoidal structures bulging above and below the disc plane. These embedded structures, predictably called bulges, come in a variety of sizes. The size, or brightness, of the bulge varies with respect to the host disc, until the bulge and disc become comparable in size and it is hard to say which structure is the host and which the guest. The ellipticals also appear to change shape gradually - or, to use the jargon, form a sequence - but this is artificial: when they are seen in the sky some look round and some quite flat, making a sequence of these projected shapes from rather flat to fully circular (see Box 1).

---

[1] At the time of publication of this article, DK was a research fellow of the European Southern Observatory in Garching, Germany, and a principal investigator of the ATLAS$^{3D}$ project (http://www-astro.physics.ox.ac.uk/atlas3d/)

The traditional classification of galaxies was built on these sequences. In 1926 Edwin Hubble, while at Mount Wilson Observatory in California and building on similar ideas to those of British astronomer John Raynolds, proposed that there are two continuous sequences of galaxies: "the elliptical nebulae and the spirals, which merge into each other." This was the basis of the famous *Hubble Sequence* (see top of Box 4), which he presented for the first time in his semi-popular 1936 book "*The Realm of Nebulae*" [1], in the form of a tuning fork (or a carving fork, for the less musically inclined). The handle of the fork is the sequence of elliptical galaxies, while the sequence of spirals is split into two prongs, depending on whether or not they have bars -- elongated rotating structures caused by instabilities in the gravitational potentials of discs. As well as the handle and the prongs, Hubble postulated a class of transition objects linking the ellipticals and spirals, and called them S0 galaxies. These were later recognised as galaxies in which one can see a disc but no spiral arms, and the bulge is as dominant a component. They are flatter than the rugby-ball-shaped ellipticals and look more like lenses, hence they are commonly called *lenticular* galaxies.

**BOX 1. Traditional galaxy types**

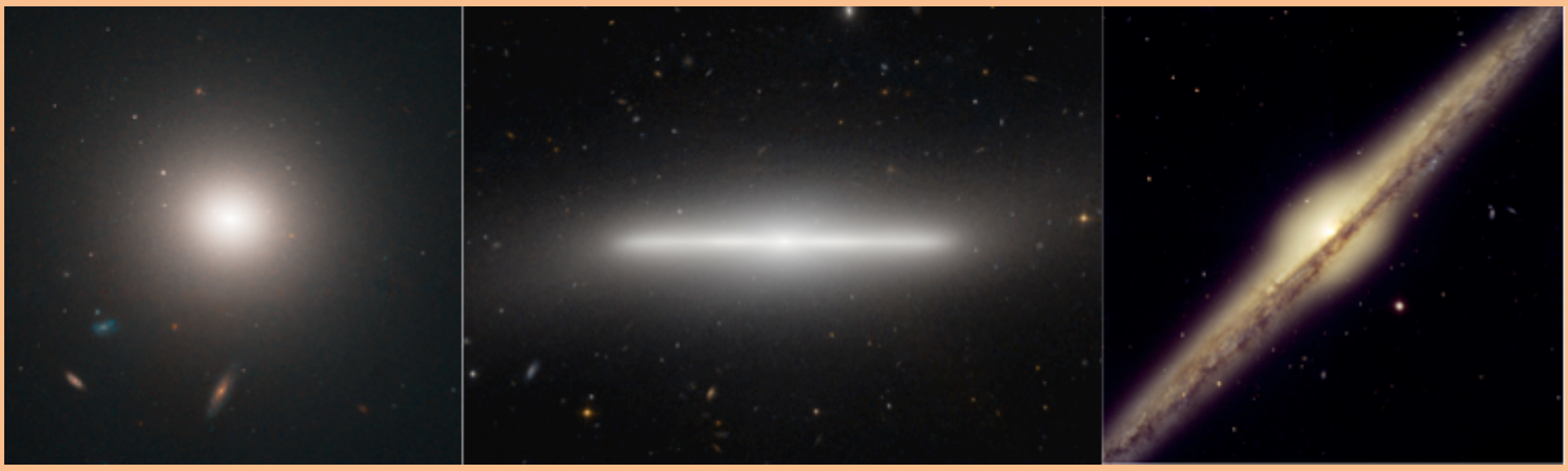

They are all fairly pleasing to the eye, but galaxies have some subtle differences and can be sorted into classes. Here are (left) a round and featureless elliptical galaxy (NGC4458), (middle) a lenticular galaxy with a thin, bright disc surrounded by a fainter halo of stars (NGC4452), and (right) a spiral galaxy with a small bulge sticking out of the centre of the disc (NGC4565). (Credit: Hubble Space Telescope/European Space Agency; European Southern Observatory; William Herschel Telescope)

Hubble designated the complex-looking galaxies in the spiral sequence as *late-type* and the featureless galaxies in the elliptical sequence of the *early-type* (later to include S0s). The misleading naming stems from an early model of stellar evolution that was later overturned with the discovery of fusion in stars (late-type stars exhibit complex spectral lines, while early-type stars have much more featureless spectra). The apparent temporal connotation is a historical left-over, and is kept, I suppose, just to confuse students.

Astronomers can make far more complicated classifications, which simultaneously explain many other features of galaxies, but the power of the Hubble Sequence is in its simplicity. It captures the fact that there are rugby balls and pancakes in the universe, that pancakes and rugby balls mix in some ways, and that this mixing is responsible for different sub-types of galaxies. There are, however, two major problems with this classification.

The first is conceptual. Although Hubble probably did not think in these terms, his fork diagram strongly suggests an evolutionary sequence, which is amplified by using the “early” and “late” designations for galaxies. Galaxy evolution is much more complex than what could be described by transformations along this simple sequence. At best, it might be possible to find individual physical properties of galaxies that change as a sequence. The second, and rather more fundamental, problem is that the sequence of ellipticals is based on a non-physical parameter: their apparent shape on the sky. In astronomy one cannot change one's perspective of the studied object, move around it and *see* its shape. (Of course, this excludes nearby bodies such as planets in our solar system and visiting asteroids.) This means that we cannot really see the shape of galaxies; we only see their projections on the sky. And a projection of a sphere, or the face of a disc, or a cigar-shaped object seen from one of its ends, will look the same: as a circle in the sky. A disc that is inclined at some moderate angle will look the same as a rugby ball from a random viewing angle. The apparent shapes of galaxies are highly degenerate - the same apparent shape can be deprojected into many different true shapes - it is virtually impossible to find out the true shape of each system just from the projection.

Astronomers, starting with Hubble himself, were aware of this problem, but lacking a good alternative it was left untouched for almost half a century. Following two decades of increasing precision in imaging, in 1996 John Kormendy, then at the University of Hawaii, and Ralf Bender, at the University Observatory in Munich, synthesised the growing evidence that there are hidden discs in some ellipticals and proposed a revision of the sequence of the early-type galaxies [2]. At some viewing angles, the shape of the contours of constant brightness, or "isophotes", reveal pointy discs rather than perfect ellipses. There are also elliptical galaxies whose isophotes show an opposite effect: their shape is more rectangular, or "boxy", as it became known (Box 2). The deviations from the perfect ellipse shape are on the order of a few percent, but easily measurable. In a classification by Kormendy and Bender, the *discy ellipticals* were linked through S0s to spirals, while the *boxy ellipticals* were separated from the rest of the continuous sequence. Their improvement was in ordering the Hubble Sequence not by the projected flatness, but by the shape of the isophotes. Unfortunately, the shapes of the isophotes are also poorly defined due to the projection effects. While the shapes of isophotes advance classification, they do not resolve the fundamental problem of galaxies being projected onto the sky.

**BOX 2. Hidden evidence**

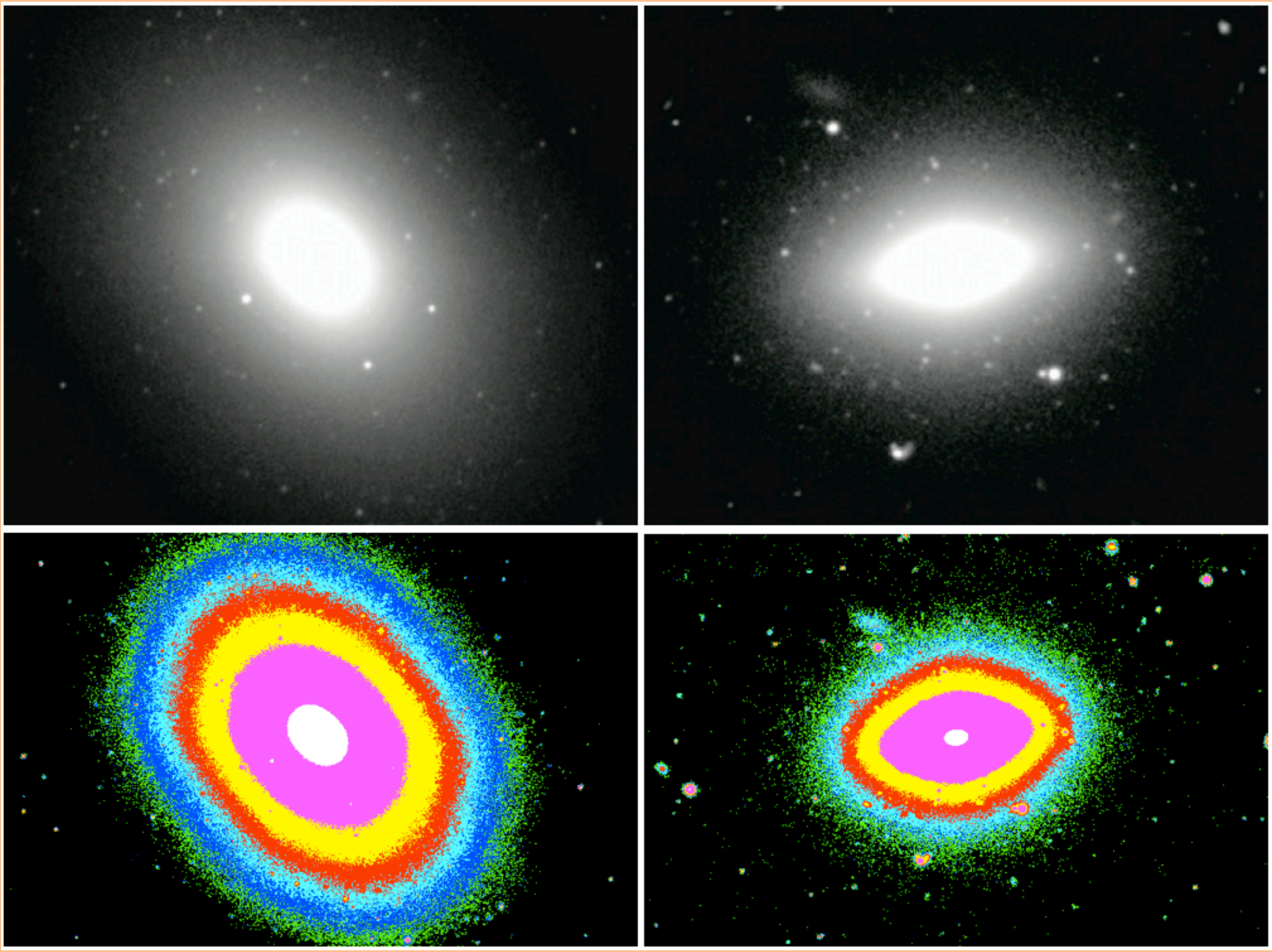

It can be hard to tell the difference between galaxies using standard optical images only (top), but when contours of constant brightness, or "isophotes", are plotted (bottom), important differences are revealed. Colours are chosen here to enhance the difference in isophote shapes between the two galaxies. Both are early types, but NGC4660 (right) has isophotes that are more pointy, or "discy" than those in NGC4365 (left), which are more rectangular, or "boxy".

## Going 3D

To see 3D in astronomy, rather than reaching for special glasses one instead needs to get used to navigating through forests of Doppler-shifted spectral lines. Spectral lines form identifiable signatures of the chemical compositions of stars, and their relative positions with respect to laboratory values -- their Doppler shifts -- reflect the speeds at which stars move. Ideally, one would like to obtain speeds of all stars in a galaxy, but there are just too many stars (usually about 100 billion) and a typical galaxy is too distant to be resolved into all its individual stars. Instead, spectra of galaxies carry statistical information about chemical and kinematical properties of stars in the observed region.

For spiral galaxies, such as our home the Milky Way, it has been known for some time that their stars move rather fast, and, to a good approximation, on circular orbits around the galactic nuclei. That is what happens in discs, anyway. For elliptical galaxies, the initial idea was that how flat they are is in some way related to the motions of their stars: flatter ellipticals have fast-moving stars and the more bulbous ellipticals have slow-moving stars. However, the first spatially resolved kinematics of galaxies showed that some of the ellipticals rotate much slower even than was expected; so slow that they could not be supported by rotation alone, but instead by a sort of random motion of stars. Their shape could be explained if they were objects with three different axes of symmetry, which allow for complex and multi-structural stellar orbits. Ellipticals became much more complicated and interesting than ever before.

The first attempts to measure kinematics of elliptical galaxies in the late 1970s and 1980s were not able to cover each galaxy fully, but only to get information along certain regions, usually along one of the galaxies' principle axes. But by the mid-1990s, a new concept matured: the integral field spectrograph (IFS), which could take spectra from the full body of a galaxy in one shot. The product of IFS observations is a 3D data cube, where two dimensions are spatial (defining the positions on the sky) and the third dimension is the wavelength detected (used along with spectral lines to define the velocities).

## SAURON: the all-seeing eye

Just before the turn of the millennium a new IFS, called SAURON (the Spectroscopic Areal Unit for Research on Optical Nebulae) [3], saw its first light behind the 4.2 m-diameter William Herschel Telescope, part of the Isaac Newton Group of Telescopes on La Palma in the Canary Islands. SAURON, built by a team of astronomers and engineers led by Roland Bacon at the Centre de Recherche Astronomique de Lyon, was used for a survey of kinematical and chemical properties of nearby early-type galaxies. The SAURON Survey [4] was led by Bacon as well as Tim De Zeeuw of Leiden University and Roger Davies of the University of Oxford. It uncovered an interesting result: using kinematic maps that cover a large fraction of a galaxy's body, you can robustly determine the specific angular momentum of stars. While this physical property still depends on viewing angle, stars in discs will, in general, rotate much faster then in ellipsoids. By measuring the angular momentum one can distinguish which galaxies are fast-rotating and disc-like and which are genuine slow-rotating ellipticals (Box 3). The SAURON Survey showed that pancakes and rugby balls can be distinguished, even when projected onto the sky.

To understand the true nature of astronomical objects researches need to observe many of them, covering the full parameter space of properties such as brightness or mass. That was the goal when an international team, led by Eric Emsellem and me at the European Southern Observatory, Michele Cappellari at Oxford and Richard McDermid at the Gemini Observatory, defined a new, much larger, project -- called ATLAS$^{3D}$ [5]. For 38 nights on the William Herschel Telescope, spread over 18 months during 2007 and 2008, we used

SAURON to observe all early-type galaxies that are visible from La Palma, within 125 million light-years and above a certain brightness threshold. The outcome was 260 data cubes containing more than half a million spectra, offering an unprecedented view of early-type galaxies.

**BOX 3. Seeing stellar speeds**

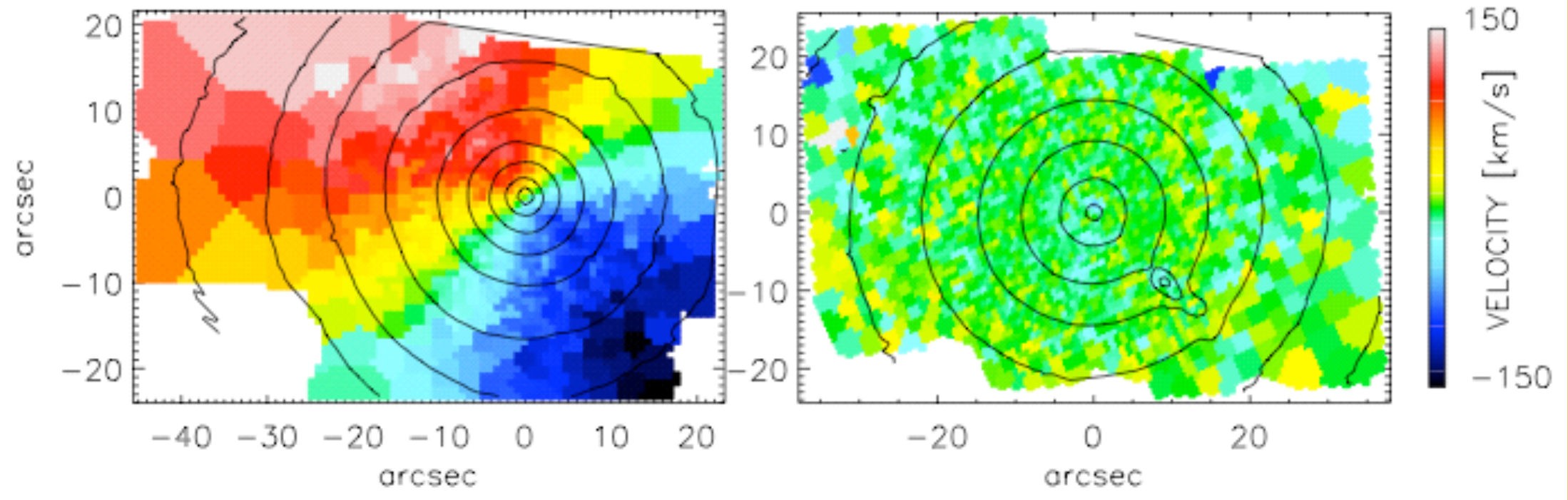

Maps of the mean stellar velocities for two early-type galaxies. Colour coding shows the mean motion of stars: regions with approaching stars are shown with blue, and regions with receding stars are red; green denotes zero mean velocity. On average, stars in the galaxy on the left show an ordered rotation around the galaxy centre, contrary to what is seen in the galaxy on the right, where stars do not exhibit a net rotation. The galaxy on the left is a fast rotator (high angular momentum) and the one on the right is a slow rotator (low angular momentum). Black lines are contours of constant brightness. Although the contours suggest that these two galaxies have very similar shapes (and belong to the same Hubble class), the kinematics clearly show their different dynamical state. The same technique and instrument were used to measure the velocity maps of 260 early-type galaxies as shown in Box 5, which shows the full complement of galaxies measured in the ATLAS$^{3D}$ survey. In this graphic the galaxies are ordered by increasing angular momentum of stars around their galactic nucleus, and the colour scale is between -200 and 200 km/s.

The first result of the ATLAS$^{3D}$ survey [6,7], presented in 2011, is that true rugby balls are very rare objects in the nearby Universe. Our data show that early-type galaxies can be separated into *fast* and *slow rotators*, and that most ellipticals are actually fast rotators and dynamically similar to spirals. They have high stellar angular momentum as expected from disc-like objects. Furthermore, the intrinsic shape of the majority of early-type galaxies is close to being axially symmetric, while their velocity maps resemble those of discs, to better than a few percent. We were surprised to see that up to 65% of galaxies traditionally classed as ellipticals are actually misclassified S0 galaxies.

All of this suggests that we need to revise the Hubble Sequence. In 1976, Sydney van den Bergh of the University of Toronto noticed that disc galaxies come in three flavours: true spirals (gas rich, making new stars), anaemic spirals (poor in gas and not making stars) and S0 galaxies (poor in gas, not making stars and not actually having spiral arms) [8]. Building on a proposal he made, the fork diagram should be replaced with a *comb* (Box 4) [9], where fast-rotating, early-type galaxies (S0 galaxies) are parallel to spirals and linked to them along the teeth of the comb, because these objects are dynamically similar and evolutionarily related. The sequence of discs, however, stops there. The handle of the comb is reserved for true, slow-rotating ellipticals, emphasising their structural and dynamical differences from all other galaxies.

## Galactic grail

So what have we learned from this game of classification? We now know that the main ingredients of the formation processes for fast and slow rotators are very different. Fast rotators are related to spirals, even though they are not pure discs of stars. The processes that shaped them, such as fly-bys or collisions with other galaxies, could have not been very severe. Instead they preserved the general symmetry and dynamical state. Fast rotators had a gentler evolution; they are likely to be spirlas that evolved largely peacefully or were involved with only a few collisions with other smaller galaxies. On the other hand, slow rotators, which do not have any relation to discs, must have lived through a much more violent past. They are the end products of fierce mergers of equal-mass galaxies and continuous bombardment with satellite galaxies, which eliminated the original disc's kinematic identity. Put simply, the new classification constrains the types and the frequency of processes that govern the formation and evolution of galaxies.

**BOX 4. Why combs are the new forks**

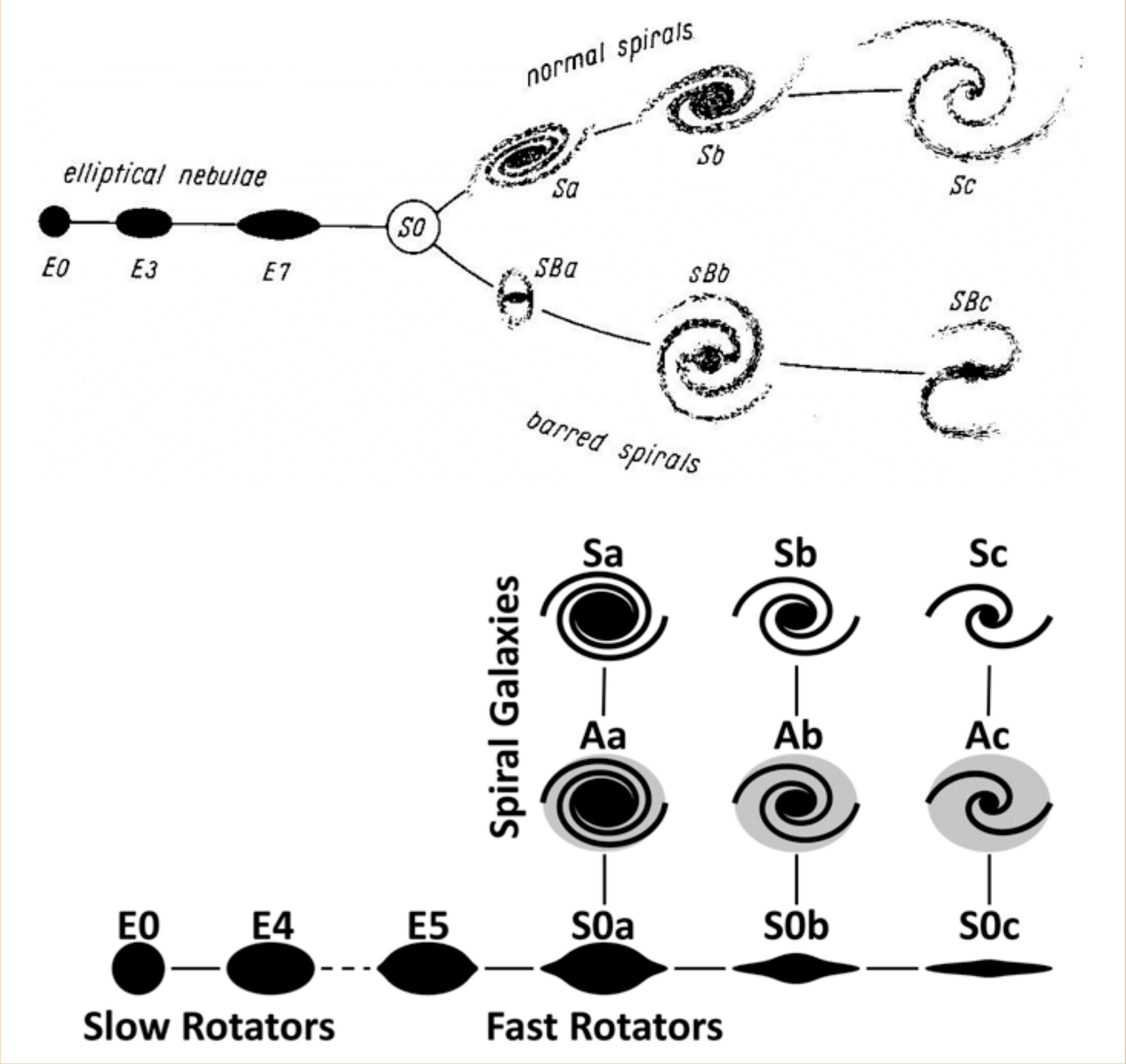

As a result of the ATLAS$^{3D}$ survey, project members have suggested that the iconic Hubble tuning fork (top) be replaced with the ATLAS$^{3D}$ Comb (bottom). Early-type galaxies are separated into slow and fast rotators, depending on their stellar angular momentum. Fast rotators, which make up the majority of early-type galaxies, are on sequence parallel to spiral galaxies (Sa-Sb-Sc), linked by anaemic spirals (Aa-Ab-Ac), which are transition objects between true spirals with large-scale spiral arms and fast rotators with no evidence for spiral structure in images. The fast rotators along the handle of the comb are traditionally classified as S0 galaxies, but they also extend into the handle of the comb, encompassing flat elliptical galaxies which often have discy isophotes. There, however, the continuous sequence of discs stops. The true ellipicals, slow rotators, occupy the rest of the handle, separated from the rest of the galaxies, indicating the different formation processes for fast and slow rotators. To emphasize their shapes, slow and fast rotators are seen edge-on, while spirals are seen top-down. (Credit: Hubble, "The Realm of Nebule", 1936, New Haven: Yale University Press; Cappellari et al. 2011, MNRAS, 413, 813)

**BOX 5. Galaxy atlas.**

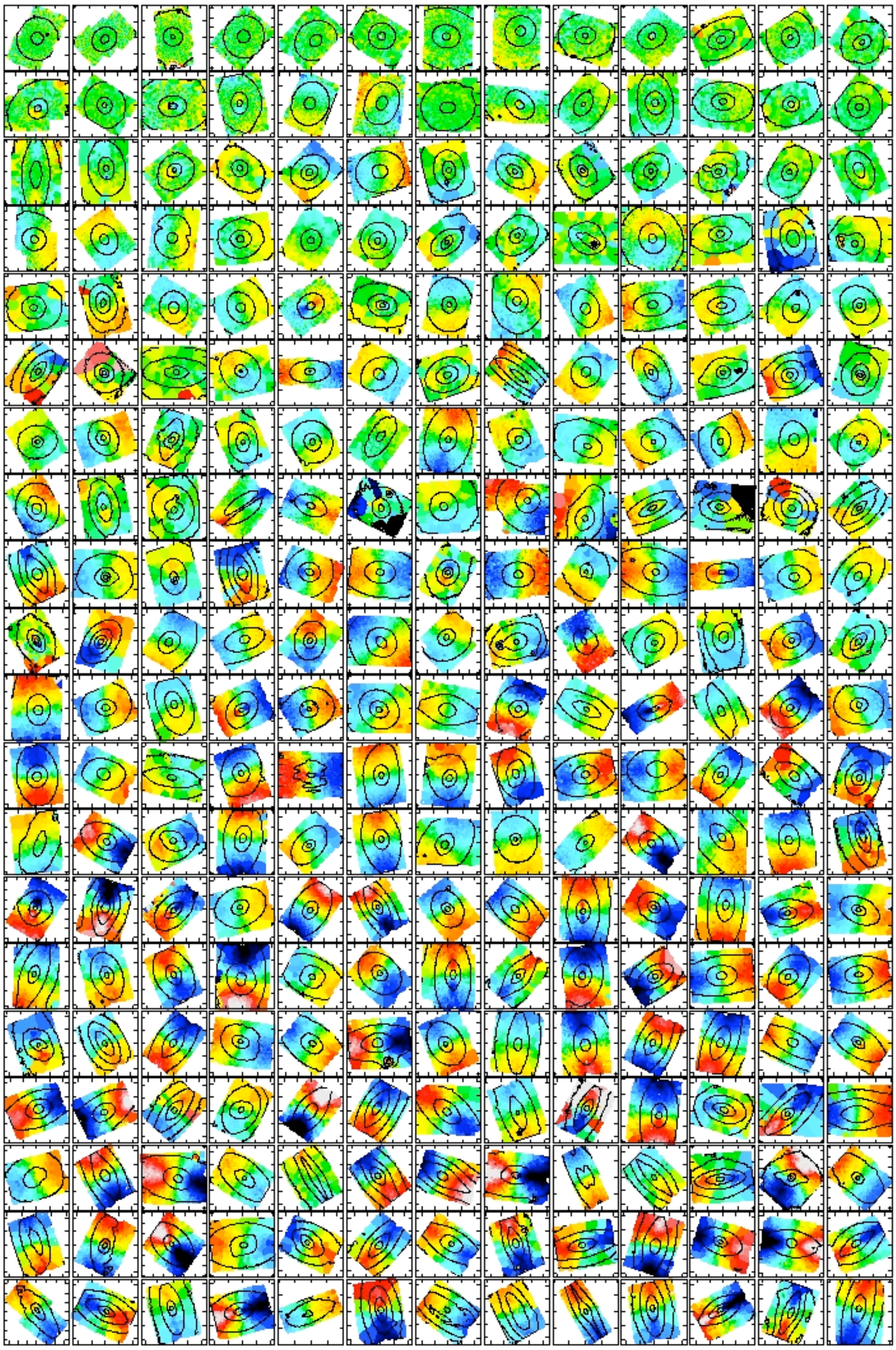

These 260 elliptical galaxies can now be visualised in 3D and their motions properly understood. (Colour coding and black lines as described in Box 3.) (Credit: Krajnović et al. 2011, MNRAS, 414, 2923)

The ATLAS$^{3D}$ Project is not only about galaxy classification. It is a multi-wavelength survey, including optical IFS observations to explore stellar kinematics and dynamics, as well as deep imaging to look for faint structures on large scales as probes of galaxy interactions. We also use radio interferometry (from the centimetre to sub-millimetre regime) to probe the atomic and molecular content of galaxies and to look for the evidence of recent or ongoing star formation. One part of the project consists of analysing these various observations, while the other is an effort in numerical simulations, where galaxies are created on computers and smashed into each other, or whole Universes are built and let to evolve to test what we have learned from observations. This work is done by an international team spanning 12 time zones. The aim of ATLAS$^{3D}$ is to set a benchmark that can be used to study the evolution of galaxies from when the Universe was half its current age or even younger. But our ultimate goal - the galactic holy grail - is to understand how galaxies form and evolve, given the sad truth that there is not much jam in Space.